\documentclass[review,12pt]{elsarticle}

\usepackage{amssymb}

\begin{document}

\begin{frontmatter}

\title{Temperature dependence of the electronic structure of A-site ordered perovskite CaCu$_3$Ti$_4$O$_{12}$: Angle-integrated and -resolved photoemission studies}




\author[1]{H. J. Im}
\ead{hojun@hirosaki-u.ac.jp}
\author[1]{T. Sakurada}
\author[2]{M. Tsunekawa}
\author[1]{T. Watanabe}
\author[3,5]{H. Miyazaki}
\author[4,5]{S. Kimura}

\address[1]{Graduate School of Science and Technology, Hirosaki University, Hirosaki 036-8224, Japan}
\address[2]{Faculty of Education, Shiga University, Otsu 520-0862, Japan}
\address[3]{Department of Physical Science and Engineering, Nagoya Institute of Technology, Nagoya 466-8555, Japan}
\address[4]{Graduate School of Frontier Biosciences, Osaka University, Suita 565-0871, Japan}
\address[5]{UVSOR Facility, Institute for Molecular Science, Okazaki 444-8585, Japan}

\begin{abstract}

We have investigated the electronic structure of A-site ordered CaCu$_3$Ti$_4$O$_{12}$ as a function of temperature by using angle-integrated and -resolved photoemission spectroscopies.
Intrinsic changes of the electronic structure have been successfully observed in the valence band region by the careful consideration of charging effects.
The obtained photoemission results have revealed that the intensity of the nearly non-dispersive Cu 3$d$-O 2$p$ hybridized bands at the binding energy of $\sim$2 eV increases with decreasing temperature from 300 to 120 K.
This suggests that the density of the localized states, caused by the strong correlation effects, enhances as temperature decreases.

\end{abstract}


\end{frontmatter}


\section{Introduction}
A-site ordered perovskite CaCu$_3$Ti$_4$O$_{12}$ (CCTO) has attracted much attention due to the colossal dielectric constant ($\varepsilon$) and the intriguing properties such as the Mott insulator phase.
The value of $\varepsilon$ is as high as 10$^4$$-$10$^5$ over the wide range of temperature (100$-$600 K) in the frequencies of 10$-$1 MHz \cite{Subr2000, Rami2000}.
This implies that CCTO is a promising material for applications such as energy storage and random access memory devices.
However, the mechanism of the colossal $\varepsilon$ has not been well established.
The unique behavior of the temperature dependence of $\varepsilon$ has been considered as a key to understand its mechanism;
$\varepsilon$ dramatically drops down to $\sim$100 around 100$-$200 K without any structural phase transition, which is quite different behavior from those of the typical ferroelectric materials such as BaTiO$_3$ \cite{Rami2000}.
Several scenarios have been proposed to explain the colossal $\varepsilon$.
One of such scenarios is the internal boundary layer capacitance (IBLC), in which the grain boundary and/or the twinning of the insulating boundary are assumed \cite{Sinc2002, Chio2004}.
This has used mainly to explain the frequency dependence of the colossal $\varepsilon$.
Another scenario is the polaron relaxation with a variable range hopping (VRH) conductivity by considering the localized states.
This has focused mainly on the understanding of the temperature dependence of the colossal $\varepsilon$ \cite{Zhang2004, Wang2007, Kroh2008}.
Besides, A-site disorder of Ca and Cu has been suggested as the origin of the colossal $\varepsilon$ \cite{Zhu2007}.

These debates require clarification of the underlying electronic structure of CCTO \cite{Home2001, Home2003, Litv2003, Im2013, Im2015, Tezu2017}.
Photoemission spectroscopy (PES) is one of the most powerful methods to directly observe the electronic structure \cite{Hufn1995, Im2008}.
Actually, we have clarified the band dispersion using angle-resolved photoemission (ARPES) measurements of the high-quality single crystalline CCTO at room temperature \cite{Im2013}.
The results have pointed out that the strong correlation effects cause the Mott insulator phase of CCTO.
Recently, the electronic structure in the unoccupied states and the band gap ($\sim$1.5$-$1.8 eV) have been also revealed in the combination of the inverse photoemission (IPES) and angle-integrated photoemission (AIPES) measurements \cite{Im2015}.
However, the PES study as a function of temperature has not been reported due to the difficulties to deal with charging effects in the measurement of insulators.

Here, we have performed the temperature-dependent AIPES and ARPES measurements on the single crystalline CCTO from 120 to 300 K.
Thanks to the careful investigation of charging effects, we could successfully observe the modification of the electronic structure with the temperature variation.
The results suggest that the intensity of the non-dispersive Cu 3$d$-O 2$p$ hybridized bands at the binding energy of $\sim$2 eV increases with decreasing temperature, indicating the enhancement of the localized states.

\section{Experimental details}
High quality CCTO single crystals were grown by the floating-zone method.
We have carried out the temperature-dependent PES experiments at the beamline 5U of UVSOR synchrotron, which has a soft X-ray monochromator, the so-called SGM-TRAIN, with entrance and exit slits \cite{Ito2007}.
ARPES and AIPES measurements have been performed in the temperature range from 120 to 300 K at the photon energy ($h \nu$) of 90 eV.
The clean surface of the (100) plane was prepared by $in$-$situ$ cleaving in the ultra-high vacuum less than $\sim$2 $\times$ $10^{-8}$ Pa.
The Fermi level ($E_F$) was referred to that of the Au thin film.
Resolution was set to be about 0.2 eV at $h \nu$ = 90 eV and the slit size of 100 $\mu$m in both entrance and exit slits.

\section{Results and Discussion}
\begin{figure*}
\begin{center}
\includegraphics[width=125mm,clip]{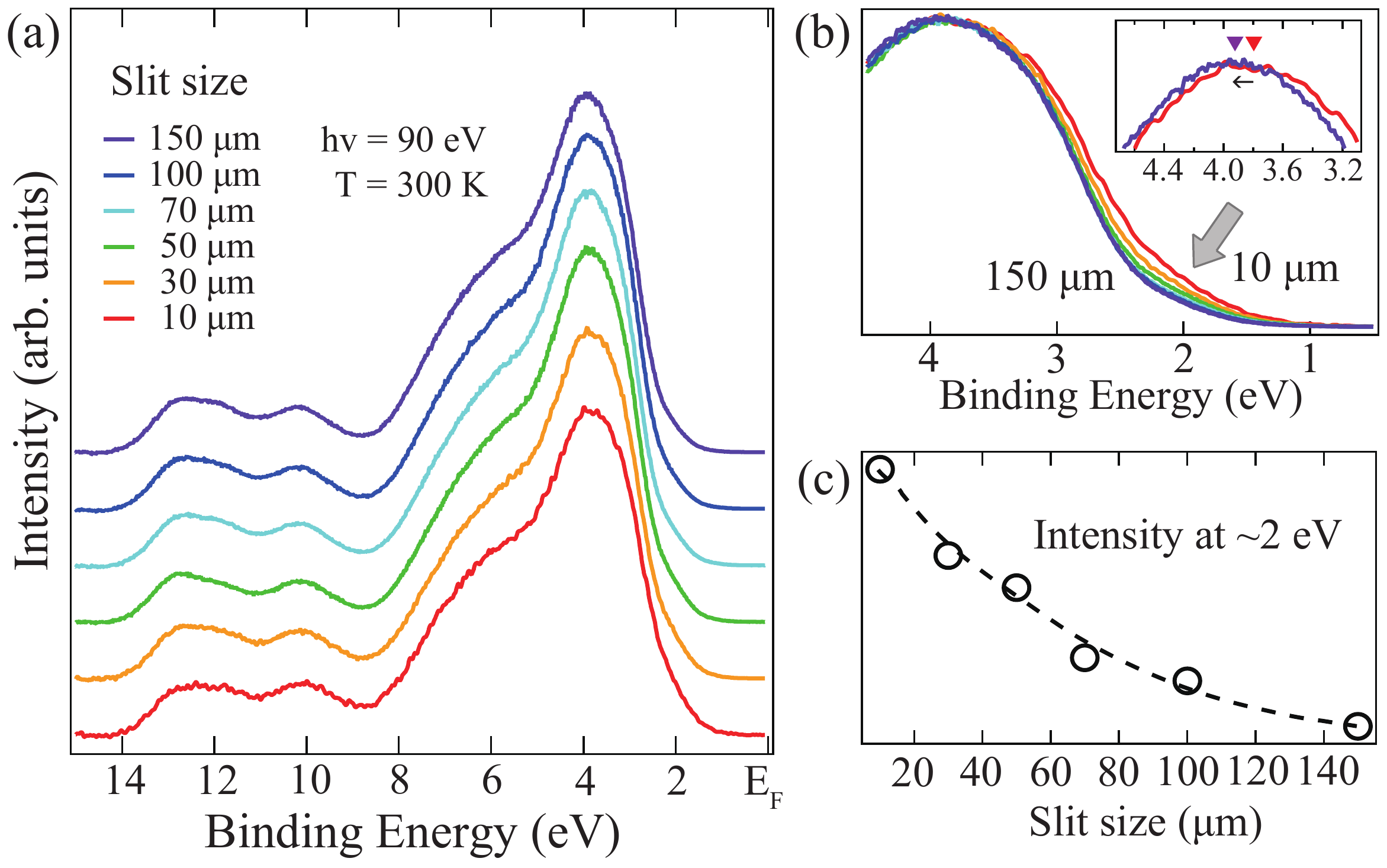}
\caption{\label{fig:figure1}
(a) AIPES spectra of CCTO with the different incident photon-flux controlled by the slit size. A set of the entrance slit of 10 $\mu$m and the exit slit of 10 $\mu$m is simply denoted by the slit size of 10 $\mu$m. (b) Enlarged AIPES spectra without offset. The inset shows the AIPES spectra of the slit sizes of 10 and 150 $\mu$m near $\sim$3.8 eV. (c) Change of the intensity at $\sim$2 eV as a function of the slit size. Dashed-line is a guide to the eye.
}
\end{center}
\end{figure*}

During the PES measurement of semiconductors or insulators, the high intensity of the incident photon-flux often gives rise to the electrical charging of samples \cite{Hufn1995, Vere1998}.
This must be properly dealt with to obtain the intrinsic PES spectra.
In order to investigate charging effects of CCTO, AIPES measurements were performed with changing both of the entrance and exit slit sizes from 10 to 150 $\mu$m, fixing the incident photon energy ($h \nu$ = 90 eV) and the temperature ($T$ = 300 K).
Photon-flux was monitored by the current of the gold-mesh located in the lower flow side of the exit slit, which changes from $\sim$0.08 to $\sim$6 nA as the slit size changes from 10 to 150 $\mu$m.
Figure 1(a) shows the AIPES spectra of CCTO in valence band region.
In all the AIPES spectra with the different slit sizes, the four typical peak features were observed in good agreement with the previous results \cite{Im2013, Im2015};
the intense Cu 3$d$ peak at $\sim$3.8 eV, the broad peak of O 2$p$ states around 5$-$8 eV, the small broad shoulder of mainly Cu 3$d$-O 2$p$ hybridized states at $\sim$2 eV, and the atomic-like satellite peaks in 9$-$14 eV.
In Fig. 1(b), the AIPES spectra were enlarged to investigate the Cu 3$d$-O 2$p$ hybridized states at $\sim$2 eV.
For the sake of comparison, the spectral intensities were normalized to the Cu 3$d$ peak at $\sim$3.8 eV.
We can clearly observe that the intensity of the Cu 3$d$-O 2$p$ hybridized states decreases with increasing the photon flux.
Figure 1(c) is the plot of its intensity variation as a function of the slit size.
Besides, the small shift of Cu 3$d$ peak was observed in the slit size of 150 $\mu$m as shown in the inset of Fig. 1(b).
Consequently, we can summarize the charging effects of CCTO during the PES experiments as following two tendencies.
One is the decrease of the Cu 3$d$-O 2$p$ hybridized states at $\sim$2 eV and the other is the shift of Cu 3$d$ peak at $\sim$3.8 eV to the higher binding energy as the incident photon-flux increases.

\begin{figure*}
\begin{center}
\includegraphics[width=125mm,clip]{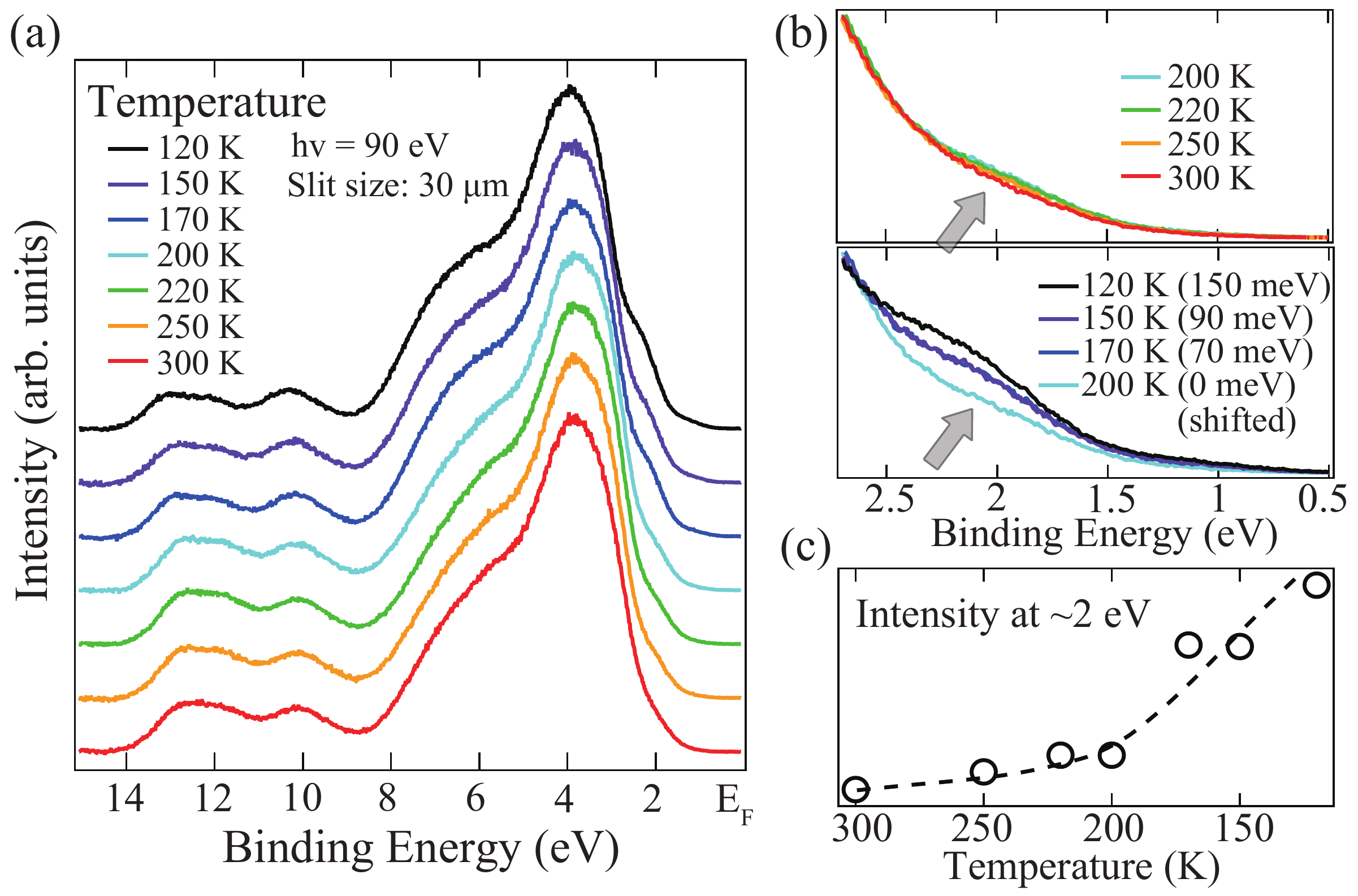}
\caption{\label{fig:figure2}
(a) Temperature dependent AIPES spectra of CCTO in the slit size of 30 $\mu$m. (b) Enlarged AIPES spectra near $\sim$2 eV in 200$-$300 K (upper panel) and 120$-$200 K (lower panel). In the lower panel, the AIPES spectra were shifted to the lower binding energy. The quantities of the energy shift were added inside the parenthesis. (c) Change of the intensity at $\sim$2 eV as a function of the temperature. Dashed-line is a guide to the eye.
}
\end{center}
\end{figure*}

Figure 2(a) shows the temperature-dependent AIPES spectra of the valence band region obtained at $h \nu$ = 90 eV and the slit size of 30 $\mu$m, where there are no change of Cu 3$d$ peak position ($\sim$3.8 eV) and the small decrease of the intensity of the Cu 3$d$-O 2$p$ hybridized bands ($\sim$2 eV) as discussed in Fig. 1.
Here, it should be mentioned that we could not use the slit size of 10 $\mu$m because of too low photon flux to carry out the temperature-dependent PES measurements within the beamtime.
In the temperature range of 200$-$300 K, the position of Cu 3$d$ peak at $\sim$3.8 eV does not change with decreasing temperature, indicating that the charging effects are negligible.
Further decreasing temperature below 200 K, the Cu 3$d$ peak is a little shifted to the higher binding energy due to the charging effects;
the shifted energies were 70, 90, and 150 meV at 170, 150, and 120 K, respectively, as depicted in the lower pannel of Fig. 2(b).
On the other hand, it is found that the intensity of Cu 3$d$-O 2$p$ hybridized bands at $\sim$2 eV increases with decreasing temperature, which is opposite behavior to that of the charging effects as discussed in Fig. 1.
This indicates that the above variation of the peak intensity does not come from the charging effects but from the intrinsic properties of CCTO.
The upper panel of Fig. 2(b) is the plot of the enlarged AIPES spectra around Cu 3$d$-O 2$p$ hybridized states in the temperature range of 200$-$300 K.
The density of the Cu 3$d$-O 2$p$ hybridized states monotonously increases with decreasing temperature.
Below 200 K, the increase of the peak intensity is more obviously observed with decreasing temperature as shown in the lower panel of Fig. 2(b).
The intensity variation was summarized in Fig. 2(c).
However, it is difficult to turn out whether the noticeable increase of the peak intensity below 200 K is intrinsic or not, because of a possibility of the influence of the charging effects.

\begin{figure*}
\begin{center}
\includegraphics[width=135mm,clip]{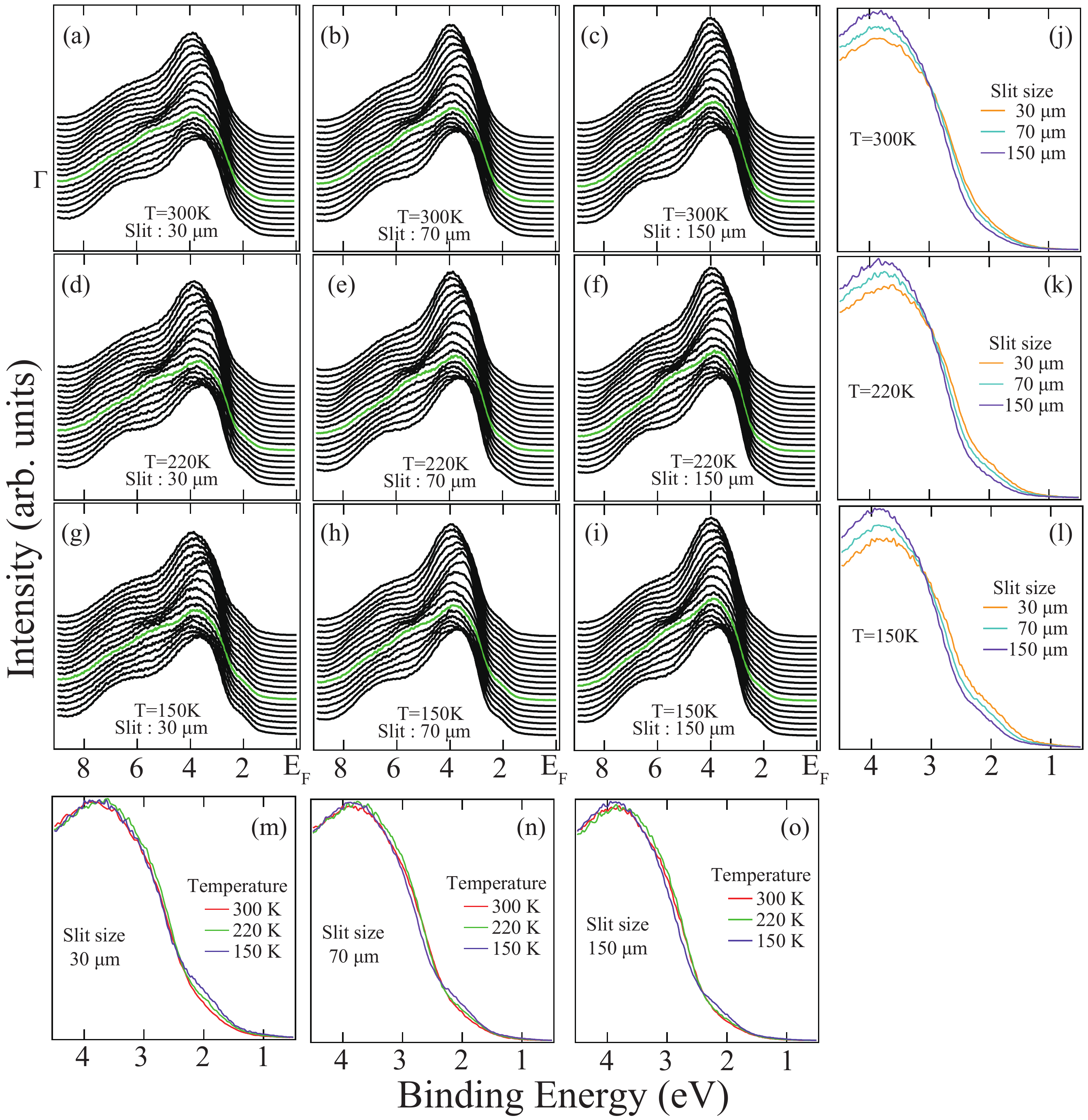}
\caption{\label{fig:figure3}
(a)$-$(i) ARPES spectra of CCTO at $T$ = 150, 220, 300 K in the slit sizes of 30, 70, 150 $\mu$m. All ARPES spectra were normalized to the integrated intensity in the valance band region from $E_F$ to 10 eV. The ARPES spectra at $T$ = 150 K were shifted to the lower binding energy by 90 meV. (j)$-$(l) ARPES spectra at $\Gamma$-point as a function of the slit size at the fixed temperatures. (m)$-$(o) ARPES spectra at $\Gamma$-point as a function of the temperature at the fixed slit sizes.
}
\end{center}
\end{figure*}

For a deeper insight of the temperature dependence of Cu 3$d$-O 2$p$ hybridized bands, we carried out the temperature dependent ARPES measurements.
Figures 3(a)$-$3(i) show the energy distribution curves (EDCs) of CCTO obtained at $h \nu$ = 90 eV in the slit sizes of 30, 70, 150 $\mu$m and the temperatures of 300, 220, 150 K.
The EDCs obtained at $T$ = 150 K were shifted to the lower binding energy by 90 meV as in the AIPES spectra (Fig. 2).
All EDSs are normalized to the integrated intensity in the valance band region from $E_F$ to 10 eV.
The green line denotes the EDC at the $\Gamma$-point.
The observed band dispersions are well consistent with the previous ARPES results;
the small dispersion of Cu 3$d$ bands at $\sim$3.8 eV, the highly dispersive O 2$p$ bands at $\sim$5$-$8 eV near the $\Gamma$-point, and the nearly non-dispersive Cu 3$d$-O 2$p$ hybridized bands at $\sim$2 eV \cite{Im2013}.
In Figs. 3(j)$-$3(l), the EDCs at the $\Gamma$-point were plotted together without offset for the sake of comparison.
We find that the decrease of the peak intensity at $\sim$2 eV with increasing the slit size is more clearly observed than those of the AIPES spectra in Fig. 2.
And, such charging effects become more obvious at the lower temperature.
In Figs. 3(m)$-$3(o), the increase of peak intensity $\sim$2 eV with decreasing temperature was clearly observed, which are the intrinsic variation of the electronic structure as revealed in the AIPES results.
It should be also noted that the Cu 3$d$-O 2$p$ hybridized bands are nearly non-dispersive and has the localized character originated from the strong correlation effects between electrons \cite{Im2013, Im2015}.
This suggests that the localized states at $\sim$2 eV, caused by the strong correlation effects, enhance with decreasing temperature.

Finally, let us discuss a relation between the obtained PES results and the colossal $\varepsilon$.
Even though we cannot determine if the large enhancement of the localized states below 200 K is intrinsic or not as discuss in the AIPES spectra (Fig. 2), it is interesting that the temperature region is roughly consistent with where the VRH conductivity has occurred \cite{Zhang2004, Wang2007}.
In addition, it should be emphasized that the scenarios to explain the colossal $\varepsilon$ in the introduction have been based on the insulating phase of CCTO, which actually originates from the strong correlation effects as evidenced by the PES studies \cite{Im2013, Im2015}.
Therefore, the strong correlation effects such as the Mott insulator phase should be taken into account for the more accurate understanding of the colossal $\varepsilon$ \cite{He2002, Rami2004, Kotl2006}.

\section{Conclusions}

We have performed the temperature-dependent AIPES and ARPES measurements on the high-quality single crystalline CCTO from 120 to 300 K.
By careful investigation of the charging effects, we could successfully observe the temperature dependence of the intrinsic electronic structure.
The AIPES and ARPES results have revealed that the intensity of the nearly non-dispersive Cu 3$d$-O 2$p$ hybridized bands at $\sim$2 eV increases with decreasing temperature, indicating the enhancement of the density of the localized states originated from the strong correlation effects.
We believe that the obtained results would provide an underlying physics to understand the origin of the colossal $\varepsilon$.

\section*{Acknowledgements}
This work was partly supported by the UVSOR Facility Program (22-525) of the Institute for Molecular Science.


\end{document}